\documentstyle[12pt,epsfig]{article} 
\begin{document}
\title{Generalized Uncertainty Principle and Quantum Electrodynamics }
\author{ A. Camacho 
\thanks{email: acamacho@janaina.uam.mx} \\
Physics Department, \\
Universidad Aut\'onoma Metropolitana-Iztapalapa. \\
P. O. Box 55-534, C. P. 09340, M\'exico, D.F., M\'exico.}
\date{}
\maketitle

\begin{abstract}
In the present work the role that a generalized uncertainty principle could play in the quantization of the electromagnetic field is analyzed. It will be shown that we may speak of a Fock space, a result that implies that the concept of photon is properly defined. Nevertheless, in this new context the creation and annihilation operators become a function of the new term that modifies the Heisenberg algebra, and hence the Hamiltonian is not anymore diagonal in the occupation number representation. Additionally, we show the changes that the energy expectation value suffers as result of the presence of an extra term in the uncertainty principle. The existence of a deformed dispersion relation is also proved. 
\end{abstract}
\bigskip
\bigskip


\section{Introduction}

As the literature shows, the quest of detectable effects stemming from the different proposals that attempt to obtain a quantum theory of gravity has grown rapidly in the last few years [1, 2]. A very interesting consequence of some of these effects comprises the possible  existence of corrections to Lorentz symmetry [2]. 
The possibilities that these kind of extensions to Lorentz symmetry could open have far--reaching implications, since they could explain three incongruities between the standard model predictions and observational results [3], namely: (i) the GZK cutoff [4]; (ii) photons, with cosmic origin, of energy 10 $TeV$, or more, should not be seen due to backgrounds induced pair production [5]; (iii) the apparent inconsistency between the predictions for the longitudinal development in extensive air showers of high enery hadronic particles, and the current experimental readouts [6]. As already claimed, all these three problems could be solved by means of a deformed dispersion relation [7].

Additionally, string theory and quantum gravity approaches suggest the existence of a minimal observable distance [8]. The possible connection between this minimal observable distance and a Generalized Uncertainty Principle (GUP) has already been pointed out [9], and this remark has defined a very interesting research area. Moreover, a modification of the Heisenberg algebra could have important physical consequences, for instance, it could imply ultraviolet regularization in field theory [10]. 

Though the presence of a GUP is also closely related to spacetime noncommutativity [11], the main weight in the analysis of the consequences upon field theory of this kind of quantum gravity effects has been concentrated upon the path integral approach (see [11] and reference therein), but of course, the study, in the context of canonical quantization, of noncommutativity of spacetime has already been consi\-dered [12], though not in a so complete manner as in the path integral case.

In the present work we will consider a GUP and introduce it in the quantization process of the electromagnetic field. This will be done starting, as usual [13], from the analogy between a harmonic oscillator and each one of the degrees of freedom of an electromagnetic field. The modification to the Heisenberg algebra will be done introducing in the commutator of the position and momentum operators an extra term, whose presence is justified by quantum gravity effects. 

It will be shown that the concept of Fock space can be conserved, this point (which could seem trivial) has not been analyzed with the required care. In this context the already studied ambiguities [14], or problems [15], with the notion of Fock space, for a quantum field theory in a curved spacetime, do not consider the existence of a GUP as a fundamental part of the formalism. We also deduce the changes in the energy expectation values that this GUP renders. In principle this fact could be detected. 

Additionally, in this new approach the Hamiltonian of the electromagnetic field is not anymore diagonal in the occupation number representation. A deformed dispersion relation will be deduced.
Finally, we will confront some of our predictions against the results of a previous work [16], where the case of a quantum harmonic oscillator (considering a GUP) was analyzed from the point of view of its energy and eigenfunctions spectra. Focusing only on the energy spectrum, it will be shown that our results do match with those of [16]. 
\bigskip

\section{Concept of Photon and GUP}
\bigskip

Let us now consider the vector potential $\vec{A}$, of an electromagnetic field, such that for the case of periodic boundary conditions we have the Fourier decomposition

{\setlength\arraycolsep{2pt}\begin{eqnarray}
\vec{A}({\vec{x}}, t) = {1\over \sqrt{V}}\sum_{\vec{k}}\sum_{\alpha = 1, 2}\Bigl(c_{\vec{k}\alpha}\hat{e}^{(\alpha)}\exp\Bigl\{i(\vec{k}\cdot\vec{x}- \omega t)\Bigr\} + c^{\ast}_{\vec{k}\alpha}\hat{e}^{(\alpha)}\exp\Bigl\{-i(\vec{k}\cdot\vec{x}- \omega t)\Bigr\}\Bigr).
\end{eqnarray}}

At this point we must comment that we employ the so--called transversality condition [13], a condition that, mathematically, reads $\nabla\cdot\vec{A} = 0$. Additionally, $\hat{e}^{(\alpha)}$ denotes the polarization direction, and $V$ is the volume where the field confined. 

Here we assume that the classical field equations are obtained in the usual way [13]. This assumption defines a starting point that is not the same as in other works, for instance [17], where the classical field equations do show from the very begining the effects of loop quantum gravity. This difference stems from the fact that in the aforementioned works any effective matter Hamiltonian is defined as the expectation value of the corresponding quantum gravity operator in a semiclassical mixed state, the one describes a flat metric and the involved matter field. In our a\-pproach we do not consider a particular model in the quantization of the gravitational field, but only analyze the consequences of GUP (which seems to emerge in many of these approaches) in the context of the modifications that this extension of the Heisenberg algebra could introduce in the definition of the creation and annihilation operators. 
Of course, a more realistic approach should consider the changes, in the Maxwell equations, that quantum gravity effects could cause. Nevertheless, this more ge\-neral idea requires, unavoidably, the use of a very particular model for quantum gravity in order to determine the extra  terms that quantum gravity renders.  Here the main goal is the deduction of effects (in the quantization of the electromagnetic field) that have a general validity.

As usual, we find that there is an analogy between the degrees of freedom of the radiation field and a set of uncoupled harmonic oscillators [13]. This point is a direct consequence of the assumption that the classical field equations suffer no mo\-difications. In other words, if the modifications, in the classical realm, that quantum gravity effects could provoke, are taken into account, then this analogy could show changes, for instance, the similarity could be, under this new scheme, between the degrees of freedom of the radiation field and a set of coupled harmonic oscillators. This scenario could be, for instance, the case if the classical field equations lose their linearity.

The Hamiltonian reads

{\setlength\arraycolsep{2pt}\begin{eqnarray}
H = {1\over 2}\sum_{\vec{k}}\sum_{\alpha = 1, 2}\Bigl(\omega^2q^2_{\vec{k}\alpha} + p^2_{\vec{k}\alpha}\Bigr),
\end{eqnarray}}

\noindent where

{\setlength\arraycolsep{2pt}\begin{eqnarray}
q_{\vec{k}\alpha} =  {1\over c}\Bigl(c_{\vec{k}\alpha} +  c^{\ast}_{\vec{k}\alpha}\Bigr),
\end{eqnarray}}

\noindent and

{\setlength\arraycolsep{2pt}\begin{eqnarray}
p_{\vec{k}\alpha} =  -{i\omega\over c}\Bigl(c_{\vec{k}\alpha} -  c^{\ast}_{\vec{k}\alpha}\Bigr).
\end{eqnarray}}
\bigskip

In the usual theory, the quantization is done considering $p_{\vec{k}\alpha}$ and $q_{\vec{k}\alpha}$ as quantum operators, such that $[q_{\vec{k}\alpha}, p_{\vec{k}'\alpha'}] = i\hbar\delta_{\vec{k}\vec{k}'}\delta_{\alpha\alpha'}$ (the remaining commutators are assumed always to vanish) [13, 18]. At this point we suppose that there is a GUP present, the one contains no mimimal uncertainty in momentum [16]

{\setlength\arraycolsep{2pt}\begin{eqnarray}
[q_{\vec{k}\alpha}, p_{\vec{k}'\alpha'}] = i\hbar\delta_{\vec{k}\vec{k}'}\delta_{\alpha\alpha'}\Bigl(\Pi + \beta p^2_{\vec{k}\alpha}\Bigr).
\end{eqnarray}}
\bigskip

Here $\beta$ is a constant, which is related to the existence of a minimal observable length [1, 2, 8, 10]. At this point one question appears in connection with this GUP, namely, how to define the Fock space? Indeed, its definition depends upon the so--called creation and annihilation operators [13], but in this new situation it can be readily seen that the usual definition of creation and annihilation operators (as a function of the position and momentum operators) can not work, since it does not lead to expression (5).

Let us now consider the following possibility, as a generalization for these two operators

{\setlength\arraycolsep{2pt}\begin{eqnarray}
a_{\vec{k}\alpha} = {1\over\sqrt{2\hbar\omega}}\Bigl(\omega q_{\vec{k}\alpha} + i[p_{\vec{k}\alpha} + f(p_{\vec{k}\alpha})]  \Bigr),
\end{eqnarray}}

{\setlength\arraycolsep{2pt}\begin{eqnarray}
a^{\dagger}_{\vec{k}\alpha} = {1\over\sqrt{2\hbar\omega}}\Bigl(\omega q_{\vec{k}\alpha} - i[p_{\vec{k}\alpha} + f(p_{\vec{k}\alpha})]  \Bigr).
\end{eqnarray}}

Here $f(p_{\vec{k}\alpha})$ is a function that satisfies three conditions, namely: (i) in the limit $\beta\rightarrow 0$ we recover the usual definition for the creation and annihilation operators, ; (ii) if $\beta \not= 0$, then we have (5), and; (iii) $[a_{\vec{k}\alpha}, a^{\dagger}_{\vec{k}'\alpha'}] = i\hbar\delta_{\vec{k}\vec{k}'}\delta_{\alpha\alpha'}$. It is readily seen that the following function satisfies the aforementioned restrictions

{\setlength\arraycolsep{2pt}\begin{eqnarray}
f(p_{\vec{k}\alpha}) = \sum_{n=1}^{\infty}{(-\beta)^n\over 2n+1}p^{2n+1}_{\vec{k}\alpha}.
\end{eqnarray}
\bigskip

Condition (iii) means that the usual results, in relation with the structure of the Fock space, are valid in our case, for instance, the definition of the occupation number operator, $N_{\vec{k}\alpha} = a^{\dagger}_{\vec{k}\alpha}a_{\vec{k}\alpha}$, the interpretation of $a^{\dagger}_{\vec{k}\alpha}$ and $a_{\vec{k}\alpha}$ as creation and annihilation operators, respectively, etc., etc. [13, 18].

Clearly, the relation between $p_{\vec{k}\alpha}$ and $a_{\vec{k}\alpha}$, $a^{\dagger}_{\vec{k}\alpha}$ is not linear, and from the Hamiltonian, expression (2), we now deduce that it is not diagonal in the occupation number representation.

In order to have a look at the consequences of GUP upon the energy spectrum, let us now consider 

{\setlength\arraycolsep{2pt}\begin{eqnarray}
f(p_{\vec{k}\alpha}) = -{\beta\over 3}p^{3}_{\vec{k}\alpha}.
\end{eqnarray}
\bigskip

Latter it will be proved that this is not a bad approximation, but at this moment this restriction allows us to find, explicitly, $p_{\vec{k}\alpha}$ as a function of $a_{\vec{k}\alpha}$ and $a^{\dagger}_{\vec{k}\alpha}$, namely

{\setlength\arraycolsep{2pt}\begin{eqnarray}
p_{\vec{k}\alpha} = -i{\sqrt{\hbar\omega\over 2}}\Bigl(a_{\vec{k}\alpha} -a^{\dagger}_{\vec{k}\alpha}\Bigr)
\Bigl[\Pi - {\sqrt{\hbar\omega\beta\over 8}}(a_{\vec{k}\alpha} -a^{\dagger}_{\vec{k}\alpha})\Bigr].
\end{eqnarray}

Clearly, if $\beta =0$ we recover the usual case [13]. Rephrasing the Hamiltonian as a function of the creation and annihilation ope\-rators we find (here we omit the term $1/2$, as is usually done in the context of quantum electrodynamics [13, 18]), then it reads

{\setlength\arraycolsep{2pt}\begin{eqnarray}
H = \sum_{\vec{k}}\sum_{\alpha = 1, 2}\hbar\omega\Bigl[N_{\vec{k}\alpha} + {\sqrt{\hbar\omega\beta\over 8}}
g(a_{\vec{k}\alpha}, a^{\dagger}_{\vec{k}\alpha}) + \beta{(\hbar\omega)^2\over 16}h(a_{\vec{k}\alpha}, a^{\dagger}_{\vec{k}\alpha})\Bigr].
\end{eqnarray}}
\bigskip

Here we have introduced two functions, $g(a_{\vec{k}\alpha}, a^{\dagger}_{\vec{k}\alpha})$ and $h(a_{\vec{k}\alpha} a^{\dagger}_{\vec{k}\alpha})$

{\setlength\arraycolsep{2pt}\begin{eqnarray}
g(a_{\vec{k}\alpha}, a^{\dagger}_{\vec{k}\alpha})= a^3_{\vec{k}\alpha}  
- N_{\vec{k}\alpha}a_{\vec{k}\alpha} - a_{\vec{k}\alpha}N_{\vec{k}\alpha} - a_{\vec{k}\alpha}\nonumber\\
 - (a^{\dagger}_{\vec{k}\alpha})^3 + N_{\vec{k}\alpha}a^{\dagger}_{\vec{k}\alpha} + a^{\dagger}_{\vec{k}\alpha}N_{\vec{k}\alpha} + a^{\dagger}_{\vec{k}\alpha},
\end{eqnarray}}
\bigskip

\noindent and

{\setlength\arraycolsep{2pt}\begin{eqnarray}
h(a_{\vec{k}\alpha}, a^{\dagger}_{\vec{k}\alpha})= a_{\vec{k}\alpha}^4 + a_{\vec{k}\alpha}^2(a^{\dagger}_{\vec{k}\alpha})^2 - a_{\vec{k}\alpha}^3a^{\dagger}_{\vec{k}\alpha}- a_{\vec{k}\alpha}^2a^{\dagger}_{\vec{k}\alpha}a_{\vec{k}\alpha} \nonumber\\
+ (a^{\dagger}_{\vec{k}\alpha})^2a_{\vec{k}\alpha}^2 + (a^{\dagger}_{\vec{k}\alpha})^4 - 
(a^{\dagger}_{\vec{k}\alpha})^2a_{\vec{k}\alpha}a^{\dagger}_{\vec{k}\alpha}- (a^{\dagger}_{\vec{k}\alpha})^3a_{\vec{k}\alpha} \nonumber\\
-  a_{\vec{k}\alpha}a^{\dagger}_{\vec{k}\alpha}a_{\vec{k}\alpha}^2-a_{\vec{k}\alpha}(a^{\dagger}_{\vec{k}\alpha})^3 + a_{\vec{k}\alpha}a^{\dagger}_{\vec{k}\alpha}a_{\vec{k}\alpha}a^{\dagger}_{\vec{k}\alpha}+ a_{\vec{k}\alpha}(a^{\dagger}_{\vec{k}\alpha})^2a_{\vec{k}\alpha} \nonumber\\
- a^{\dagger}_{\vec{k}\alpha}a_{\vec{k}\alpha}^3 - 
a^{\dagger}_{\vec{k}\alpha}a_{\vec{k}\alpha}(a^{\dagger}_{\vec{k}\alpha})^2+ a^{\dagger}_{\vec{k}\alpha}a^2_{\vec{k}\alpha}a^{\dagger}_{\vec{k}\alpha} + a^{\dagger}_{\vec{k}\alpha}a_{\vec{k}\alpha}a^{\dagger}_{\vec{k}\alpha}a_{\vec{k}\alpha}.
\end{eqnarray}}
\bigskip

\section{Conclusions}
\bigskip

Let us consider the energy expectation value for the one--photon situation, whose frequency is $\omega$. Employing our last expression we find that 

{\setlength\arraycolsep{2pt}\begin{eqnarray}
<1\vert H\vert 1> = \hbar\omega\Bigl[1 + {13\over 16}\hbar\omega\beta\Bigr].
\end{eqnarray}}
\bigskip

The momentum $\vec{P}$ associated to the electromagnetic field is

{\setlength\arraycolsep{2pt}\begin{eqnarray}
\vec{P} = {1\over c}\int\Bigl(\vec{E}\times\vec{B}\Bigr)d^3x = \sum_{\vec{k}}\sum_{\alpha = 1, 2}\hbar\vec{k}N_{\vec{k}\alpha}.
\end{eqnarray}}

This result can be better understood if we remember that one of our starting conditions was the validity of the usual commutators for $a_{\vec{k}\alpha}$ and $a^{\dagger}_{\vec{k}'\alpha'}$, a fact that renders (15) [13, 18]. In other words, the energy in this case changes, nevertheless, the momentum suffers no modifications at all.

In this last expression, once again, we omitted the term $1/2$. If $ck = \omega$, then we obtain an approximate expression for the dispersion relation (here $\tilde{E} = <1\vert H\vert 1>$)

{\setlength\arraycolsep{2pt}\begin{eqnarray}
\tilde{E}^2 - c^2\vec{P}^2 = {13\over 8}\beta(\hbar\omega)^3\Bigl(1 + {13\over 2^5}\beta\hbar\omega\Bigr).
\end{eqnarray}}

And in consequence, the energy becomes, approximately

{\setlength\arraycolsep{2pt}\begin{eqnarray}
\tilde{E}^2 = c^2P^2\Bigl[1 + {13\over 8}\beta cP + ({13\over 16}\beta)^2c^2P^2\Bigr].
\end{eqnarray}}

In a previous work [16] the energy spectrum of a harmonic oscillator (with GUP) was found.  The energy for the case $n = 1$ (see equation (69) in (16)), up to linear order in $\beta$, has the following form $E_1 = \hbar\omega\Bigl(1 + \beta\hbar\omega\Bigr)$.

Comparing with our energy expectation value for the one--photon situation, expression (14), we see that they match, at least in the power of the leading term in $\beta$. Moreover, we may assert that the discrepancy (${13\over 16}$ instead of $1$) stems from the fact that we have introduced an approximation for $f(p_{\vec{k}\alpha})$, expression (9), but that this restriction renders the value of the $n = 1$ state with an estimation of 80 percent.

Summing up, we have considered a GUP, and introduced it in the quantization process of the electromagnetic field. From the very outset, we have assumed that the classical field equations suffer no modifications. This condition can be understood as the roughest approximation, in the classical realm, that we could use. Clearly, a more realistic approach should consider the changes, in the Maxwell equations, that quantum gravity effects could cause, nevertheless, this idea would require the use of a very particular model for quantum gravity, and here the main goal has been the deduction of effects (in the quantization process of the electromagnetic field) that could have a general validity, independent from the employed model of quantum gravity.

It was shown that the idea of Fock space can be conserved, and in consequence we have proved that, under these conditions, the concept of photon has physical meaning. At this point we must mention the analysis of the concept of n--particle states that in connection with canonical non--perturbative quantum general relativity has recently appeared [19, 20]. In these two last references the states that could be considered n--particle states have been identified, and though [19] is a more general approach, than the one here described, its conclusions require the semiclassical limit of canonical non--perturbative quantum general relativity (see equations (4.49) and (4.50) in [20]). 

There is, in this model (equation (5.10) in [20]), a deformed dispersion relation, which may be rewritten in the form $\omega_\pm(k) = \vert k\vert\sqrt{A \pm B k}$, where $A$ and $B$ are defined by the semiclassical states. In our results, the deformed dispersion relation emerges from the effects of quantum gravity in the definition of creation and annihilation operators, expression (8), by means of GUP. Notice that, in contrast to [20], we have assumed $\omega = ck$.

Additionally, in this new approach the Hamiltonian of the electromagnetic field is not anymore diagonal in the occupation number representation. Moreover we assert that if there is a GUP (stemming either from string theory, or from loop quantum theory, etc., etc.) then the consequences, upon the quantization of the radiation field (if the concept of n--particle state is to be valid under the presence of a GUP), imply that, if $\beta > 0$, then the energy becomes larger, with respect to the case that appears in connection with the usual Heisenberg algebra (or smaller if $\beta < 0$). In principle this fact could be detected. Finally, we confronted some of our predictions against the results of a previous work, and proved that these two different approaches do coincide.

\bigskip
\bigskip
\Large{\bf Acknowledgments.}\normalsize
\bigskip

The author would like to thank A. A. Cuevas--Sosa for his help. 

\end{document}